\documentclass[twocolumn]{article}

\usepackage{arxiv}

\usepackage[utf8]{inputenc} 
\usepackage[T1]{fontenc}    
\usepackage{hyperref}       
\usepackage{url}            
\usepackage{booktabs}       
\usepackage{amsfonts}       
\usepackage{nicefrac}       
\usepackage{microtype}      
\usepackage{doi}
\usepackage{lipsum}
\usepackage{makecell}
\usepackage{amsmath}
\usepackage[numbers]{natbib}
\usepackage[capitalise,noabbrev]{cleveref}
\usepackage{graphicx}
\newcommand{\qty}[2]{{#1} {#2}}

\graphicspath{ {./images/} }
\usepackage{xcolor}
\usepackage[acronym]{glossaries}
\usepackage{fancyhdr} 
\glsdisablehyper  

\newcommand{\affit}[1]{$^{\mathrm{\textnormal{\textit{#1}}}}$}

\newacronym{ecmwf}{ECMWF}{European Centre for Medium-Range Weather Forecasts}
\newacronym{ddms}{DDMs}{Data-driven models}
\newacronym{ddm}{DDM}{data-driven model}
\newacronym{aifs}{AIFS}{Artificial Intelligence Forecasting System}
\newacronym{metno}{MET Norway}{Norwegian Meteorological Institute}
\newacronym{metcoop}{MetCoOp}{Meteorological Co-operation on Operational \acrshort{nwp}}
\newacronym{meps}{MEPS}{MetCoOp Ensemble Prediction System}
\newacronym{nwp}{NWP}{Numerical Weather Prediction}
\newacronym{synop}{SYNOP}{surface synoptic observations}
\newacronym{nmhses}{NMHSes}{national meteorological and hydrological services}
\newacronym{gnns}{GNNs}{graph neural networks}
\newacronym{gnn}{GNN}{graph neural network}
\newacronym{lam}{LAM}{limited area model}
\newacronym{gpus}{GPUs}{graphical processing units}
\newacronym{gpu}{GPU}{graphical processing unit}
\newacronym{t}{T}{2-meter temperature}
\newacronym{ws}{WS}{10-meter wind speed}
\newacronym{p6h}{P6h}{6-hour precipitation accumulation}
\newacronym{tmin}{Tmin}{daily minimum 2-meter temperature}
\newacronym{tmax}{Tmax}{daily maximum 2-meter temperature}
\newacronym{wsmax}{WSmax}{daily maximum wind speed}
\newacronym{p24h}{P24h}{24-hour precipitation accumulation}
\newacronym{mslp}{MSLP}{mean sea level pressure}
\newacronym{ifs}{IFS}{Integrated Forecast System}
\newacronym{mse}{MSE}{mean squared error}
\newacronym{rmse}{RMSE}{root mean square error}
\newacronym{ets}{ETS}{equitable threat score}
\newacronym{mlp}{MLP}{multilayer perceptron}
\newacronym{nlp}{NLP}{Natural Language Processing}

\title{Regional data-driven weather modeling with a global stretched-grid}

\author{
 Thomas Nils Nipen \affit{a*} \\ 
   \And
 Håvard Homleid Haugen \affit{a*} \\  
  \And
 Magnus Sikora Ingstad \affit{a*} \\ 
  \And
 Even Marius Nordhagen \affit{a*} \\ 
  \And
 Aram Farhad Shafiq Salihi \affit{a*} \\ 
  \And
 Paulina Tedesco \affit{a*} \\ 
  \And
 Ivar Ambjørn Seierstad \affit{a*} \\ 
  \And
 Jørn Kristiansen \affit{a} \\ 
  \And
  Simon Lang \affit{b} \\
  \And
  Mihai Alexe \affit{c} \\
  \And
  Jesper Dramsch \affit{c} \\
  \And
  Baudouin Raoult \affit{b} \\
  \And  
  Gert Mertes \affit{b} \\
  \And
  Matthew Chantry \affit{b} \\
  }

\begin{document}

\twocolumn[
\begin{@twocolumnfalse}
\maketitle
\bigskip
\begin{abstract}

A \acrfull{ddm} suitable for regional weather forecasting applications is presented. The model extends the \acrlong{aifs} by introducing a stretched-grid architecture that dedicates higher resolution over a regional area of interest and maintains a lower resolution elsewhere on the globe. The model is based on \acrlong{gnns}, which naturally affords arbitrary multi-resolution grid configurations.

The model is applied to short-range weather prediction for the Nordics, producing forecasts at \qty{2.5}{km} spatial and \qty{6}{h} temporal resolution. The model is pre-trained on 43 years of global ERA5 data at \qty{31}{km} resolution and is further refined using 3.3 years of \qty{2.5}{km} resolution operational analyses from the \acrfull{meps}. The performance of the model is evaluated using surface observations from measurement stations across Norway and is compared to short-range weather forecasts from \acrshort{meps}. The \acrshort{ddm} outperforms both the control run and the ensemble mean of \acrshort{meps} for \qty{2}{m} temperature. The model also produces competitive precipitation and wind speed forecasts, but is shown to underestimate extreme events.

\end{abstract}

\bigskip
\keywords{data-driven modeling \and weather forecasting \and graph neural networks}
\vspace{1.5cm}
\end{@twocolumnfalse}
]

\footnotetext[1]{Norwegian Meteorological Institute, Oslo, Norway}
\footnotetext[2]{ECMWF, Reading, UK}
\footnotetext[3]{ECMWF, Bonn, Germany}
\footnotetext{$^*$Equal contribution}

\section{Background}
\acrfull{ddms} are rapidly emerging as an alternative to \acrfull{nwp} models. Global \acrshort{ddm}s have been shown to produce weather forecasts that are competitive in accuracy with state-of-the-art global \acrshort{nwp} models \citep{bouallegue2024}. Unlike \acrshort{nwp} models that simulate the weather by solving physical equations, \acrshort{ddms} are typically based on artificial neural networks with parameters that are optimized by training on historical reanalyses. Although training costs are significant, \acrshort{ddms} can be several orders of magnitude cheaper to run than \acrshort{nwp} models of similar resolution, due to the efficient mapping of neural networks to \acrfull{gpus} and their ability to use a much longer timestep when integrating forward in time.

This progress has been made possible by the availability of free and reliable datasets like ERA5 \citep{hersbach2020}, which provides a long archive of consistent reanalyses at approximately \qty{31}{km} horizontal grid resolution. Several global \acrshort{ddms} have been trained using ERA5, including FourCastNet \citep{pathak2022}, Pangu-Weather \citep{bi2023}, GraphCast \citep{lam2023}, FuXi \citep{chen2023}, and the \acrfull{aifs} \citep{lang2024}.

\begin{figure*}[t]
    \centering
    \includegraphics[width=15cm]{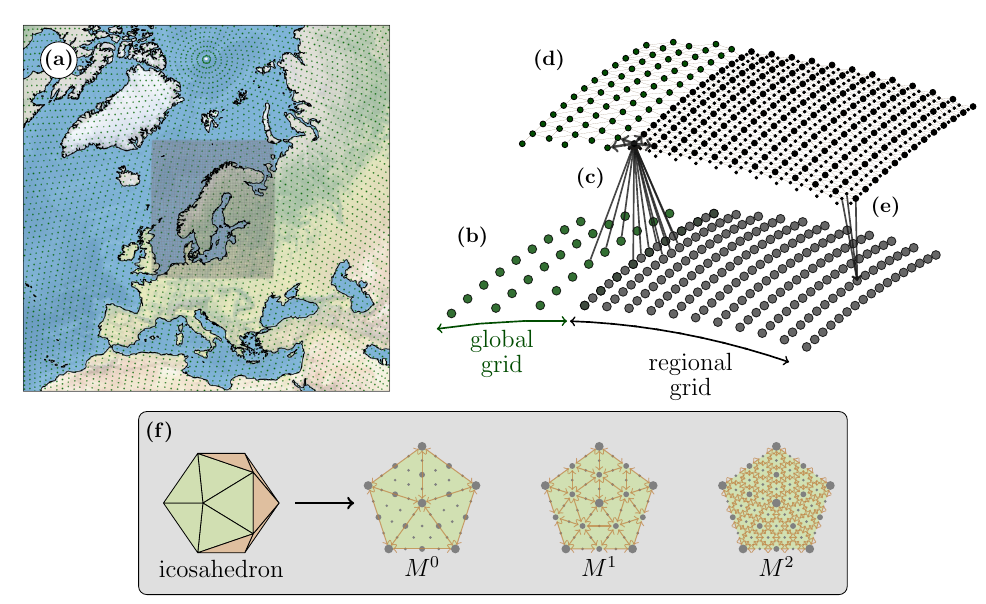}
    \caption{(a) Map with annotated grid points centered around the Nordics. Global grid points are green, regional grid points are gray. (b) Input grid on the boundary between global and regional domain. (c) The encoder processes information into a mesh node from the 12 nearest grid points. (d) The processor mesh contains latent information, and has finer refinement over the regional than the global domain. (e) The decoder processes information back to a grid node from the 3 nearest mesh nodes. (f) On the mesh, information is processed between nodes at various refinements, here represented by the three largest ($M^0, M^1, M^2$) refinement layers. }
    \label{fig:model}
\end{figure*}

The success of global \acrshort{ddms} is driving efforts to develop \acrshort{ddms} for regional purposes. In their mission to save lives and protect property, \acrfull{nmhses}, such as the \acrfull{metno}, currently rely on high-resolution regional \acrshort{nwp} models to deliver real-time weather forecasts to the public and weather-affected industries.

Additionally, many \acrshort{nmhses} produce regional reanalyses, with consistent high-resolution data spanning long time periods. These high-resolution datasets, often at a kilometer scale, offer an enormous potential for training regional \acrshort{ddms}. Due to their high resolution, these datasets provide additional information at finer spatial scales than ERA5 can represent.

Regional data-driven modeling efforts have so far focused on driving a \acrfull{lam} with initial conditions from a regional analysis and lateral boundaries from an external global model \citep{oskarsson2023, xu2024a}. This is analogous to the common separation of regional and global models in \acrshort{nwp}. 

This article presents a different approach, where a global model is developed with higher resolution over the region of interest and lower resolution elsewhere. The model is initialized from both global and high-resolution regional analyses. The goal of this stretched-grid approach is for weather systems to move seamlessly from the global domain into the regional domain and out again without any explicit treatment of the boundary between the domains. The model is based on \acrfull{gnns} \citep{scarselli2008}, which is also the architecture used in GraphCast and \acrshort{aifs}, and was introduced to weather prediction by \citep{keisler2022}. This flexible architecture allows arbitrary grids to be used, supporting our need to have variable resolution across the globe.

We target the data-driven model for use by \acrshort{metno}'s public weather forecasting app Yr (\url{https://www.yr.no}). Given Norway's intricate topography and coastline, the model must be capable of providing high-resolution forecasts. Currently, short-range weather forecasts on Yr for the Nordics are provided by the MetCoOp Ensemble Prediction System (MEPS) \citep{frogner2019} with a horizontal resolution of \qty{2.5}{km}. The performance of the model is therefore compared to the operational MEPS and the evaluation of the model is done with a focus on the needs of our end-users.    

This article is organized as follows: Section~\ref{sec:model} describes the stretched-grid approach to regional data-driven modeling. Section~\ref{sec:training} describes the hyper-parameters we chose and how the model was trained. Section~\ref{sec:evaluation} evaluates the data-driven model against current operational \acrshort{nwp} systems. Section~\ref{sec:conclusions} draws conclusions and presents future work.

\section{Stretched-grid model}
\label{sec:model}

\acrshort{gnn}-based models, such as GraphCast and \acrshort{aifs}, use an encoder-processor-decoder architecture (\cref{fig:model}). Weather parameters are provided to the model on an \textit{input grid} (\cref{fig:model}a,b). Typically, this includes data on multiple pressure levels for the current time and one or more recent times in the past. This data is then encoded into a compact latent representation of the state of the atmosphere (\cref{fig:model}c). The latent space is typically on a lower resolution grid, which we will call the \textit{processor mesh} (\cref{fig:model}d). This latent state is advanced forward one time step by the \textit{processor}. The updated state is then decoded back to the weather parameters on the input grid (\cref{fig:model}e). This process can be repeated to create forecasts of arbitrary length, which is called \textit{rollout}.

Graphs are central building blocks of \acrshort{gnns}, consisting of a set of nodes connected by edges. Nodes are used to represent the points in the input grid and in the processor mesh. As the nodes in the graphs are represented by a one-dimensional vector (i.e. is not required to be on a regular grid), arbitrary grid configurations can be defined. Graphs are well suited for a problem with variable resolution because the structural information in the model is encoded in the model features while the model weights do not depend explicitly on the shape of the graph. This allows us to dedicate a finer grid spacing in parts of the domain, which we call a \textit{stretched grid}. Although the stretched-grid approach can be used with any number of different resolution domains, we focus here on applications with a single global and a single regional domain.

Graphs are used by the encoder, processor, and decoder. These components update the nodes they are connected to by processing information from any node connected by the edges. This update is called \textit{message passing}. To aggregate the information from the contributing nodes, we use a graph transformer, which is a generalization of the transformer architecture \citep{vaswani2017} to graphs, where message passing is done via a multi-head attention mechanism \citep{dwivedi2020, shi2020}, the same GNN architecture used in the encoder and decoder of AIFS \citep{lang2024}. Our motivation for choosing the graph transformer over a pure transformer is that the graph transformer explicitly embeds directional and positional information via the edge features into the calculation of attention scores, which can be important for a multi-resolution model. 

The stretched-grid model is built using the Anemoi framework first created by \acrshort{ecmwf} and now co-developed by several meteorological organisations across Europe, including \acrshort{metno} (\url{https://anemoi.ecmwf.int/}).

\subsection{Grid and processor mesh design}

The input grid is created to match the exact grid points from the two input models. Grid points from the global model that overlap with the regional model are removed, which is a configuration referred to as \textit{cutout}. To construct the processor mesh, we follow \citep{lam2023} and start with an initial regular icosahedron with 20 faces, creating a graph with 12 nodes and 30 edges. This icosahedron is gradually refined by adding nodes and edges that divide each triangular face into four smaller triangles (\cref{fig:model}f). When a layer of refinement is added to the mesh, we keep the edges of the original lower-resolution mesh so that different levels of refinement facilitate communication on different length scales in the final multi-layer mesh. We add extra levels of refinement over the regional domain, with the aim of keeping the number of edges from the input grid to each mesh node constant between the regional and global domains.

The encoder maps data on the input grid to a hidden representation that encodes the structural information in the graph. The first step is the source node embedding, which consists of a linear layer that transforms the raw features of the source nodes (in the grid) into a latent space representation. Similarly, the destination node embedding transforms the raw features from the destination nodes (in the mesh) to a hidden dimension. Layer normalization is applied to both source and destination node embeddings. The embedded source and destination nodes, along with processed edge attributes, are passed through a graph transformer block to perform the core processing. Each processor mesh node is connected to the nearest 12 input grid points (\cref{fig:model}c). 

\subsection{Processor}
The processor evolves the latent space representation of the data on the processor mesh through a series of message-passing steps. Each message-passing step is done through a graph transformer block where every node receives information from all one-hop neighbors in the multi-layer mesh. Consequently, communication on different mesh resolutions happens simultaneously \citep{lam2023} with spatial relations embedded in the multi-head attention mechanism. In total, the processor consists of 16 such message-passing steps, in contrast to one in the encoder and decoder.

\subsection{Decoder}
The decoder maps the latent space representation back to weather parameters on the input grid. To achieve this, the embedding process performed by the encoder is reversed by a graph transformer block, followed by embedding of the input features via a skip connection. Each point in the output grid receives data from its three nearest processor mesh nodes (\cref{fig:model}e).

\section{Model training}
\label{sec:training}

\begin{table*}
 \caption{List of variables used as input and output in the model. Pressure fields are available on 50 hPa, 100 hPa, 150 hPa, 200 hPa, 250 hPa, 300 hPa, 400 hPa, 500 hPa, 700 hPa, 800 hPa, 850 hPa, 925 hPa, 1000 hPa levels. * indicates output variables only. Forcing fields are input fields only.\\}
  \centering
  \begin{tabular}{lll}
    \toprule
    Pressure level fields     & Single level fields     & Forcing fields \\
    \midrule
Geopotential height & Skin/sea-surface temperature    & Solar insolation\\
Temperature         & \qty{2}{m} temperature                  & Sine of Julian day\\
Specific humidity   & \qty{2}{m} dew point temperature        & Sine of latitude\\
U-component of wind & \qty{10}{m} u-component of wind         & Sine of local time\\
V-component of wind & \qty{10}{m} v-component of wind         & Sine of longitude\\
& Mean sea level pressure                             & Cosine of Julian day\\
& Surface air pressure                                & Cosine of latitude\\
& Surface geopotential height                         & Cosine of local time\\
& Total column integrated water                       & Cosine of longitude\\
& 6-hour accumulated precipitation*                   & Land area fraction\\
    \bottomrule
  \end{tabular}
  \label{tab:table_variables}
\end{table*}

\subsection{Data}
To train the stretched-grid model, we use both global and regional NWP analyses. As global data, we use ERA5 reanalysis \cite{hersbach2020} on its native \qty{31}{km} resolution reduced Gaussian grid for the period 1979--2022. We also use the operational analyses of the \acrfull{ifs} \citep{ifs}, interpolated to the ERA5 resolution for the time period 2020--2024, as this is what the model will use as input when run operationally.

For the regional domain, we use the analysis of the \acrfull{meps} \cite{frogner2019,muller2017} with \qty{2.5}{km} horizontal grid spacing, which covers Norway, Sweden, Finland, Denmark, and the Baltic countries. As the model underwent substantial changes, including a change to the grid configuration, on February 5, 2020, we only use analyses after this time.

\cref{tab:table_variables} summarizes the input and output variables we used from these datasets.

When performing hyper-parameter exploration, we used any data available in the time period 1979-01-01 to 2022-05-31 for training, reserving 2022-06-01 to 2023-05-31 for validation to help select the best model. We reserved a separate period for independent testing of the best model for 2023-06-01 to 2024-05-31, with results presented in \cref{sec:evaluation}.

\subsection{Training configuration}
\label{sec:training_configuration}

Many of the settings used for the model follow \cite{lang2024}, including two input time steps, the choice of loss function (squared error), per-variable weights, and optimizer. We used 1024 channels in the encoder, processor, and decoder, resulting in a total of 246 million trainable parameters.

 \begin{figure*}[t]
    \centering
    \includegraphics[width=\linewidth]{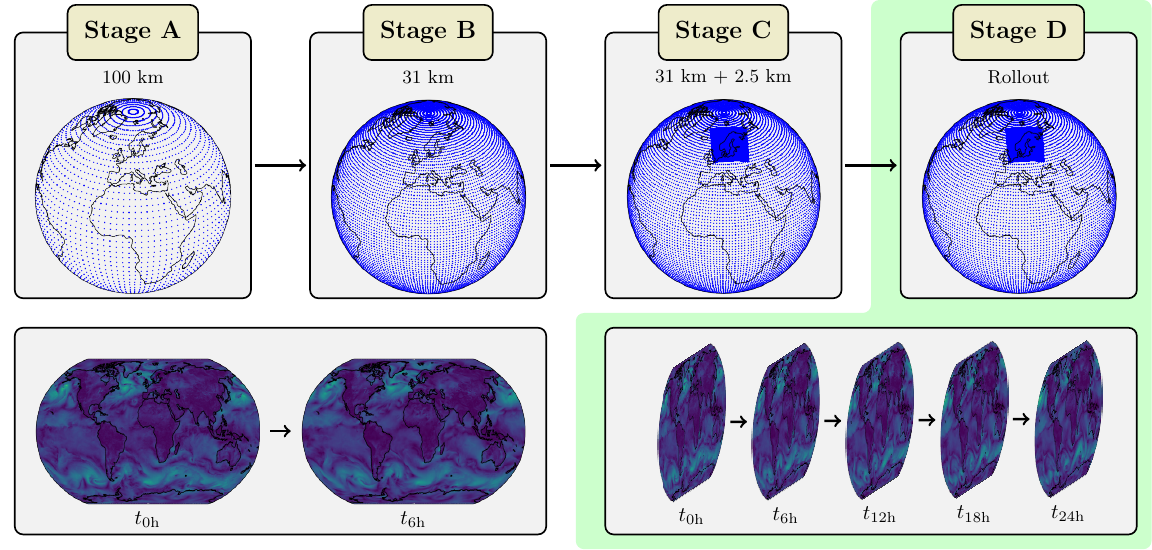}
    \caption{Model training follows a four-stage procedure. First, the \acrshort{ddm} is pre-trained on 43 years of ERA5 data with a global resolution of \qty{100}{km} (stage A) and \qty{31}{km} (stage B). In stage C, we combine the \qty{31}{km} global \acrshort{ifs} dataset with the \qty{2.5}{km} regional \acrshort{meps} dataset with a training period of 3.3 years. Finally, the model is fine-tuned by auto-regressive rollout training over four prediction time steps (stage D).}
    \label{fig:stages}
\end{figure*}

Early results showed that training a stretched-grid model using only 2-3 years of data leads to forecasts with poor synoptic developments. Therefore, we pre-trained the model on the 43 years of ERA5 data, before fine-tuning on the combined datasets. A consequence of this is that we cannot use any learnable features (\cite{lang2024}), as these are graph-specific.

We used a four-stage training procedure, as illustrated in \cref{fig:stages} and with full details in \cref{tab:table_steps}. After each stage, the scheduler was reset with a smaller starting learning rate to prevent catastrophic forgetting.

In stage A, the model was pre-trained on ERA5 upsampled to 100 km resolution. This is a relatively cheap procedure that subsequently speeds up the training of Stage B, where we trained on ERA5's full resolution.

In Stage C, we switched to using IFS upsampled to the native ERA5 resolution and added MEPS at \qty{2.5}{km} resolution in a stretched-grid configuration. In \cite{lang2024}, each output grid point's contribution to the loss is made proportional to the area that the point covers. We found that increasing the contribution of the grid points in the regional domain improved verification scores for the regional domain. In particular, we let the regional domain contribute 33\% to the total loss value despite it covering only 1.2\% of the earth's surface.

Finally, in Stage D we perform a 24 h auto-regressive rollout training to improve scores over longer lead times. This is done by incrementally increasing the rollout from two to four \qty{6}{h} time steps with 100 iterations each. Here, the scheduler is reset for each rollout step, reaching its starting learning rate after 10 warm-up iterations. We calculate the loss aggregated across the rollout steps, as in \cite{keisler2022}.

\begin{table*}[t]
\caption{The sequential training stages of the model. The learning rate is the starting learning rate after warm-up and is presented in a per-batch manner. This can be multiplied by the batch size to get the effective learning rate.\\}
\centering
\begin{tabular}{@{}lllll@{}}
\toprule
& Stage A & Stage B & Stage C & Stage D \\
\midrule
Global dataset & ERA5 & ERA5 & IFS & IFS \\
Global resolution & \qty{100}{km} & \qty{31}{km} & \qty{31}{km} & \qty{31}{km} \\
Global mesh refinement & 5 & 7 & 7 & 7 \\
Regional dataset & -- & -- & MEPS & MEPS \\
Regional resolution & -- & -- & \qty{2.5}{km} & \qty{2.5}{km} \\
Regional mesh refinement & -- & -- & 10 & 10 \\
Grid points & 40 k & 540 k & 1.5 M & 1.5 M \\
Mesh points & 10 k & 164 k & 284 k & 284 k \\
Total number of edges & 325 k & 4.9 M & 10.3 M & 10.3 M \\
Rollout & 1 & 1 & 1 & 2-4 \\
Training period (years) & 43 & 43 & 3.3 & 3.3 \\
Epochs & 101 & 10 & 39 & 3 \\
Iterations & 200 k & 15 k & 5.9 k & 300 \\
Learning rate ($\times10^{-5}$) & 6.25 & 1 & 0.4 & 0.3 \\
Warm up iterations & 1,000 & 1,000 & 1,000 & 10 \\
Batch size & 32 & 32 & 32 & 16 \\
GPU Nodes & 4 & 32 & 32 & 16 \\
GPU memory per model instance& 16 GB & 240 GB & 370 GB & 500 GB \\
Total GPU-hours & 1,750 & 2,500 & 1,900 & 180 \\
\bottomrule
\end{tabular}
\label{tab:table_steps}
\end{table*}

\subsection{Hardware}

The \acrshort{ddm} was trained on compute nodes equipped with four AMD Instinct MI250X \acrshort{gpu}s, each with 128 GB memory. To speed up training, we used data parallelism across 32 model instances for stages A--C and 16 for stage D. For stages B--D, we used model parallelism to run one instance of the model on each node. This splits the nodes and edges of the graphs across multiple GPUs and is necessary for fitting the \qty{31}{km} global and \qty{2.5}{km} regional stretched-grid model within the available \acrshort{gpu} memory. The model parallel approach is described in \cite{lang2024}.

\section{Evaluation}
\label{sec:evaluation}

This section aims to evaluate the quality of the forecasts from the perspective of a public weather forecast provider. A majority of the forecast parameters we provide on our public weather forecasting app Yr, are deterministic. The key properties we seek for our forecasts are accuracy, by low squared error, and that the frequency distribution of the forecasts match that of the observations. The latter is important to ensure that aggregated statistic over longer time periods (e.g. maximum daily wind speed) is accurate, and are important parameters for users that are not sensitive to the exact timing of weather events. As these properties are often in conflict with each other, we evaluate both separately.

To evaluate the \acrshort{ddm} from \cref{sec:training} against competing \acrshort{nwp} systems, we reserved a separate test period that we did not use when selecting the model configuration. The test period spans the time period from June 1, 2023 to May 31, 2024. The final model in Section~\ref{sec:training} was retrained on the combined training and validation period from Feb 6, 2020 to May 31, 2023 in order to increase the training period.

\subsection{Baseline NWP models}

The \acrshort{ddm} is evaluated against state-of-the-art \acrshort{nwp} models used at MET Norway. Operationally, MET Norway relies on \acrshort{meps} to provide high-resolution short-range weather forecasts for lead times up to 61 hours. This system is used for both automated weather forecasts for the general public and public weather warnings issued by duty forecasters. Additionally, they are used by downstream users such as energy, transportation, and agriculture. We extracted the control run and the ensemble mean from this modelling system.

We also include the control run from \acrshort{ifs} at 0.1$^{\circ}$ resolution, as this is a model we use to assess the added value of high-resolution models. 

\subsection{SYNOP observation network}

The models are evaluated against observations from MET Norway's network of  \acrshort{synop} stations. To allow for a fair comparison of models at different resolutions, the gridded data is interpolated bilinearly to the station locations. For temperature, both the temperature and the model terrain height are interpolated to the station point. The temperature is then adjusted based on a lapse rate of 6.5 $^{\circ}$C/km applied to the difference between the station elevation and the bilinearly interpolated model terrain height.

\subsection{Overall evaluation}

The parameters we investigate are \acrlong{t}, \acrlong{ws}, and \acrlong{p6h}. Additionally, we are interested in \qty{24}{h} aggregations, including daily minimum and maximum temperature, daily maximum wind speed, and \qty{24}{h} accumulated precipitation. Such aggregated values are used to summarize a day and are heavily used by our users. We also look at \acrfull{mslp} as a diagnostic for the model's ability to capture general synoptic developments.

\begin{figure*}[t]
    \centering
    \includegraphics[width=\linewidth]{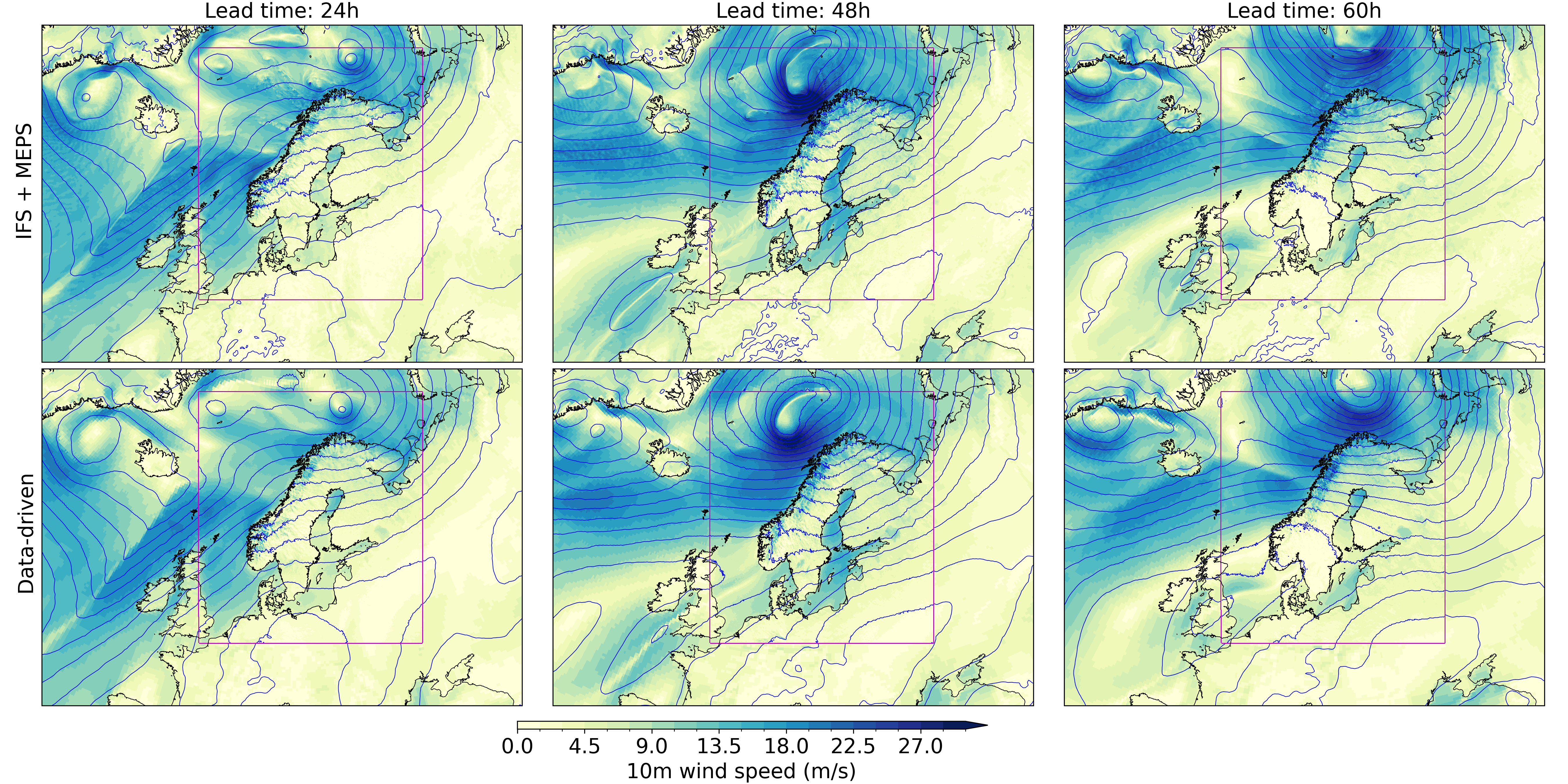}
    \caption{Forecasts of \qty{10}{m} wind speed (colormap) and mean sea-level pressure (blue isobars) produced by MEPS and IFS (top row) and the stretched-grid \acrshort{ddm} (bottom row). Forecasts were initialized at 06Z on January 27th, 2024.}
    \label{fig:forecast_time_series}
\end{figure*}

\begin{figure*}[t]
    \centering
    \includegraphics[width=\linewidth]{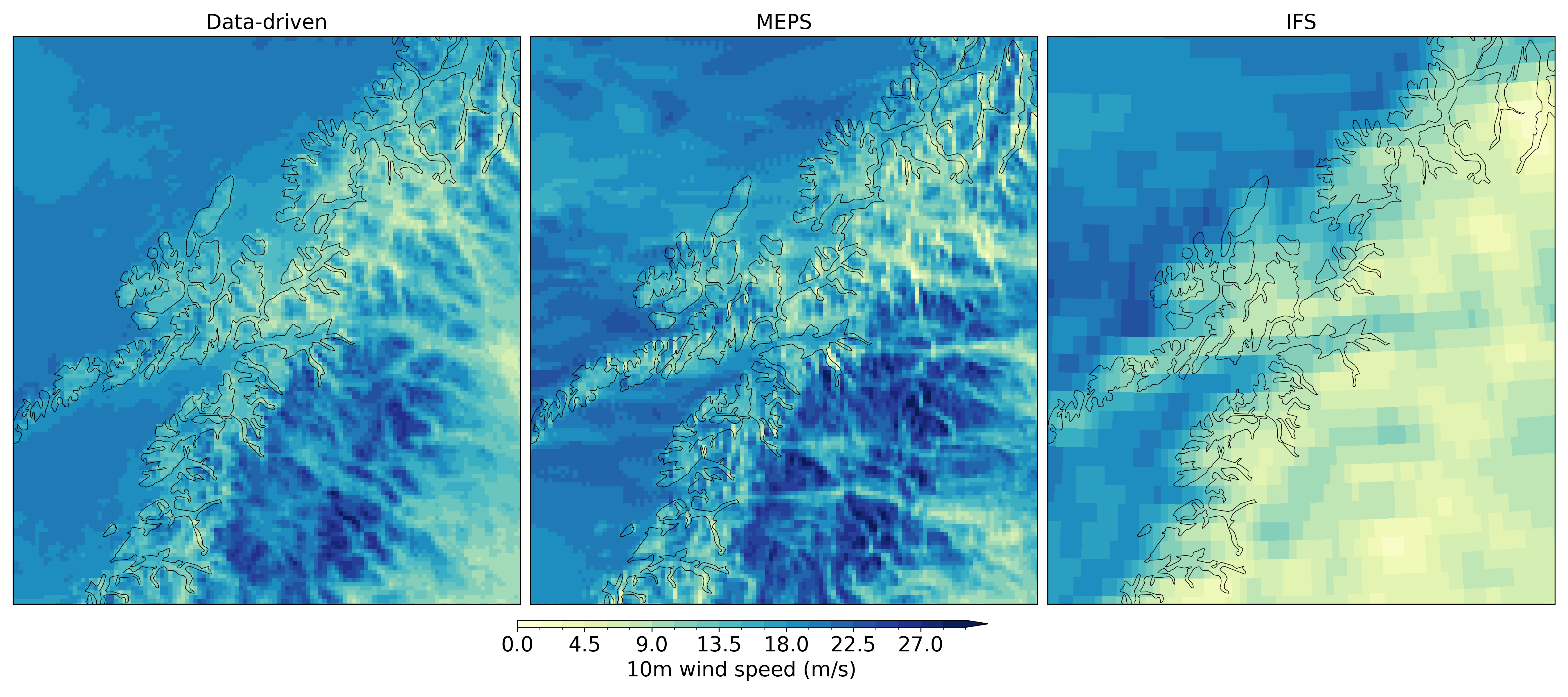}
    \caption{Forecasts of \qty{10}{m} wind speed produced by the stretched-grid model, MEPS and IFS over the mountains in northern Norway. Shows lead time \qty{24}{h} for a forecast initialized at 18Z on January 28th, 2024.}
    \label{fig:wind_mountains_case}
\end{figure*}

\cref{fig:forecast_time_series} demonstrates how the stretched grid approach is able to seamlessly move weather systems from the global into the limited area domain. This case shows the storm Ingunn for a forecast initialized at 06Z on January 27th, 2024. Ingunn can clearly be seen to move into and develop within the limited area domain. Overall, the DDM appears in this case to have a similar capability to the current operational IFS and MEPS system in simulating large-scale features of wind speed and \acrshort{mslp}. A notable difference is the slight spatial smoothing of the data-driven fields. 

\cref{fig:wind_mountains_case} shows a forecast from the same period zoomed in on the mountains in northern Norway. Like MEPS, the DDM also simulates strong winds on the mountain tops and weaker winds in the valleys. This is in contrast to the coarse resolution IFS, which fails entirely to represent the stronger winds in the mountains. This indicates that the DDM, despite some spatial smoothing, is able to represent many of the systematic wind speed features in mountainous areas.

\begin{figure}
    \centering
    \includegraphics[width=0.5\textwidth]{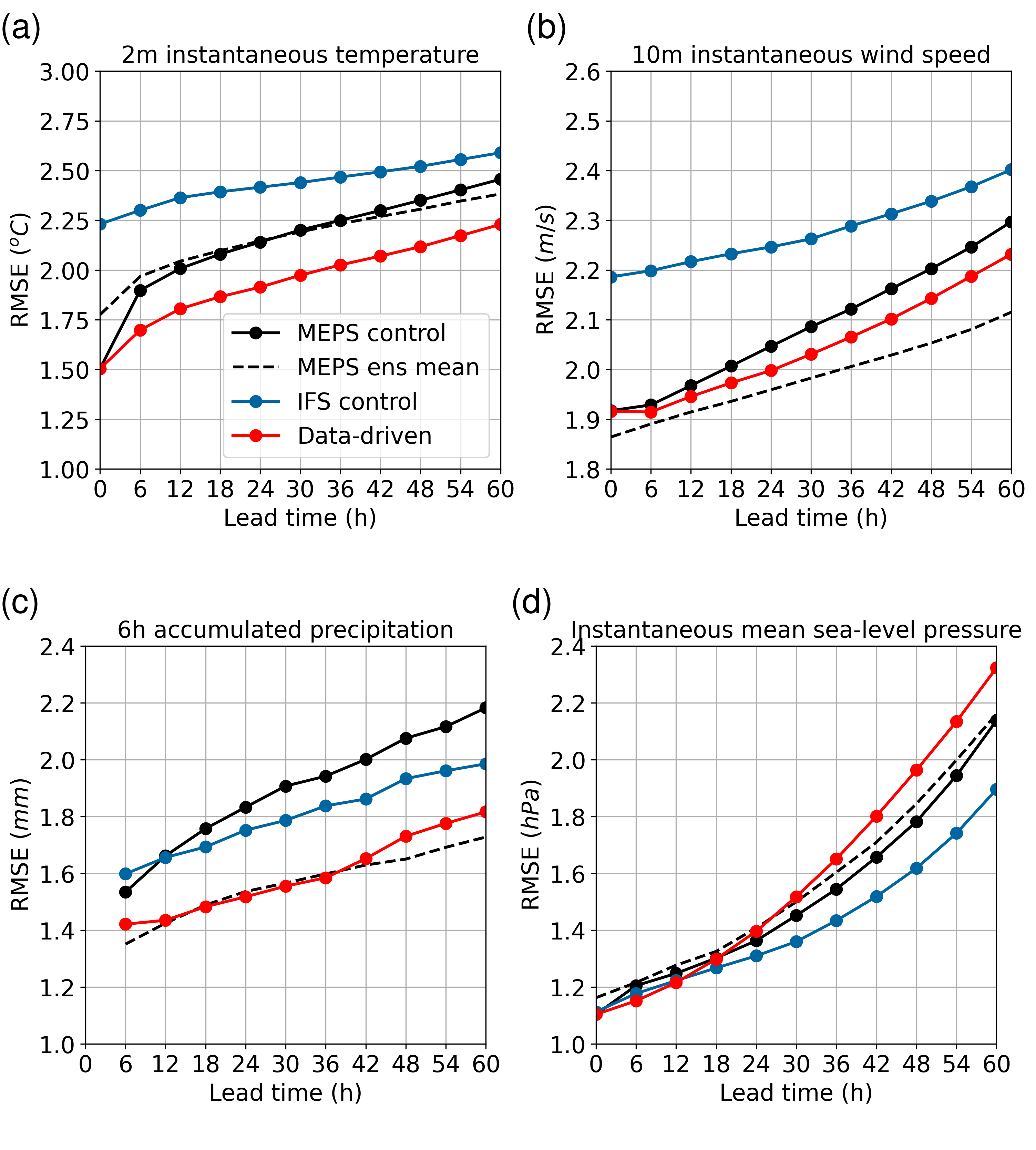}
    \caption{Root mean squared error for \qty{2}{m} temperature (a), \qty{10}{m} wind speed (b), \qty{6}{h} precipitation amount (c) and mean sea-level pressure (d). The lead time for \qty{6}{h} precipitation represents the end of the time period.}
    \label{fig:rmse}
\end{figure}

In terms of \acrfull{rmse}, the \acrshort{ddm} outperforms both \acrshort{nwp} systems for temperature (\cref{fig:rmse}a) and \qty{6}{h} precipitation (\cref{fig:rmse}c). For wind speed, the model performs similar to the MEPS control, but worse than the MEPS ensemble mean (\cref{fig:rmse}b). This is due to the fact that the \acrshort{ddm} has been trained on analyses from the control run, which tries to represent meteorological features that are unpredictable. The ensemble mean is significantly smoother and therefore scores better for all lead times. For \acrshort{mslp}, the \acrshort{ddm} is comparable to both \acrshort{nwp} systems for lead times below 24 hours (\cref{fig:rmse}d). However, the error growth is larger, and it is outperformed by both \acrshort{nwp} systems for longer lead times. We found that \acrshort{mslp} is sensitive to the area weighting of the regional domain, with a lower weight giving better \acrshort{mslp} scores but worse for the other parameters. This could indicate some over-fitting, which we will investigate further. 

\subsection{Temperature}

\acrlong{t} is greatly affected by the complex topography and coastline of Scandinavia. This makes it a challenging parameter to predict.

For instantaneous temperatures, the \acrshort{ddm} outperforms \acrshort{meps} for RMSE (\cref{fig:rmse}a). The \acrshort{ddm} improves the \acrshort{rmse} by around 24 hours compared to \acrshort{meps} control and ensemble mean. Improvements of this magnitude usually require many years of model development. A major contribution to this improvement likely comes from the way \acrshort{meps} assimilates surface temperature observations and carries that state over from the analysis into the forecast. This is evident in the large increase in error from lead time \qty{0}{h} to lead time \qty{6}{h}. \acrshort{meps} assimilates the same observations we use to verify the forecast with. This leads to a low RMSE score for lead time \qty{0}{h}. However, a large part of this is lost already at lead time \qty{6}{h}. The \acrshort{ddm} does not suffer from this problem because it is trained against analyses that have these observations assimilated and is better able to propagate information from the assimilated state forward in time.

\begin{figure}
    \centering
    \includegraphics[width=0.5\textwidth]{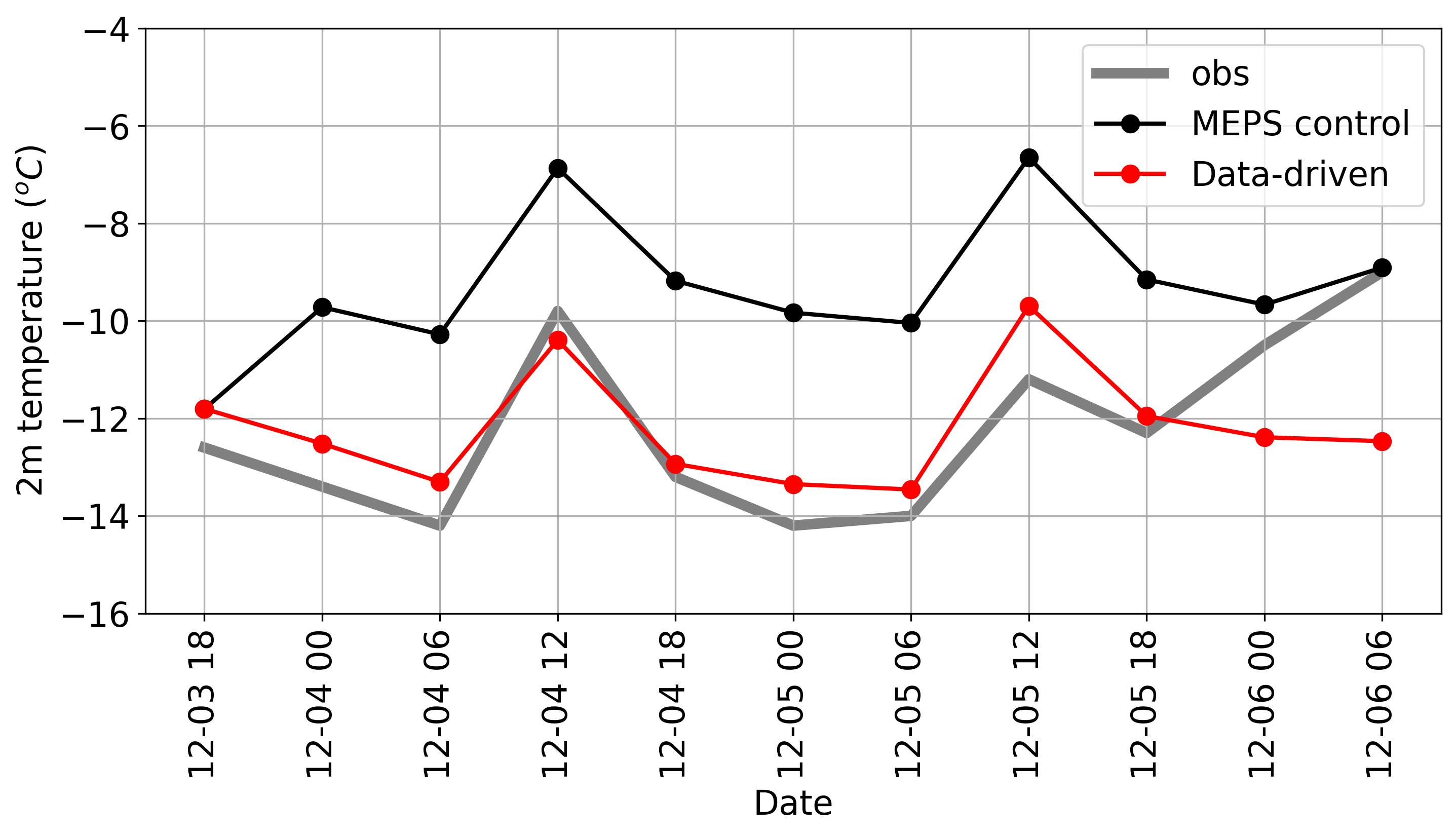}
    \caption{\qty{2}{m} temperature time series for the Blindern station for the forecast initialized at 18Z on December 3rd, 2023.}
    \label{fig:timeseries}
\end{figure}

As an example, \cref{fig:timeseries} shows that when the analysis increment is large, and the NWP model is not able to carry the increment forward, a large systematic error is present for the remainder of the forecast. Note that we verify the forecasts using the same observations employed in the data assimilation. This implies that we cannot expect to get the same forecast improvements over areas where no observations are assimilated. 

\begin{figure}
    \centering
    \includegraphics[width=0.5\textwidth]{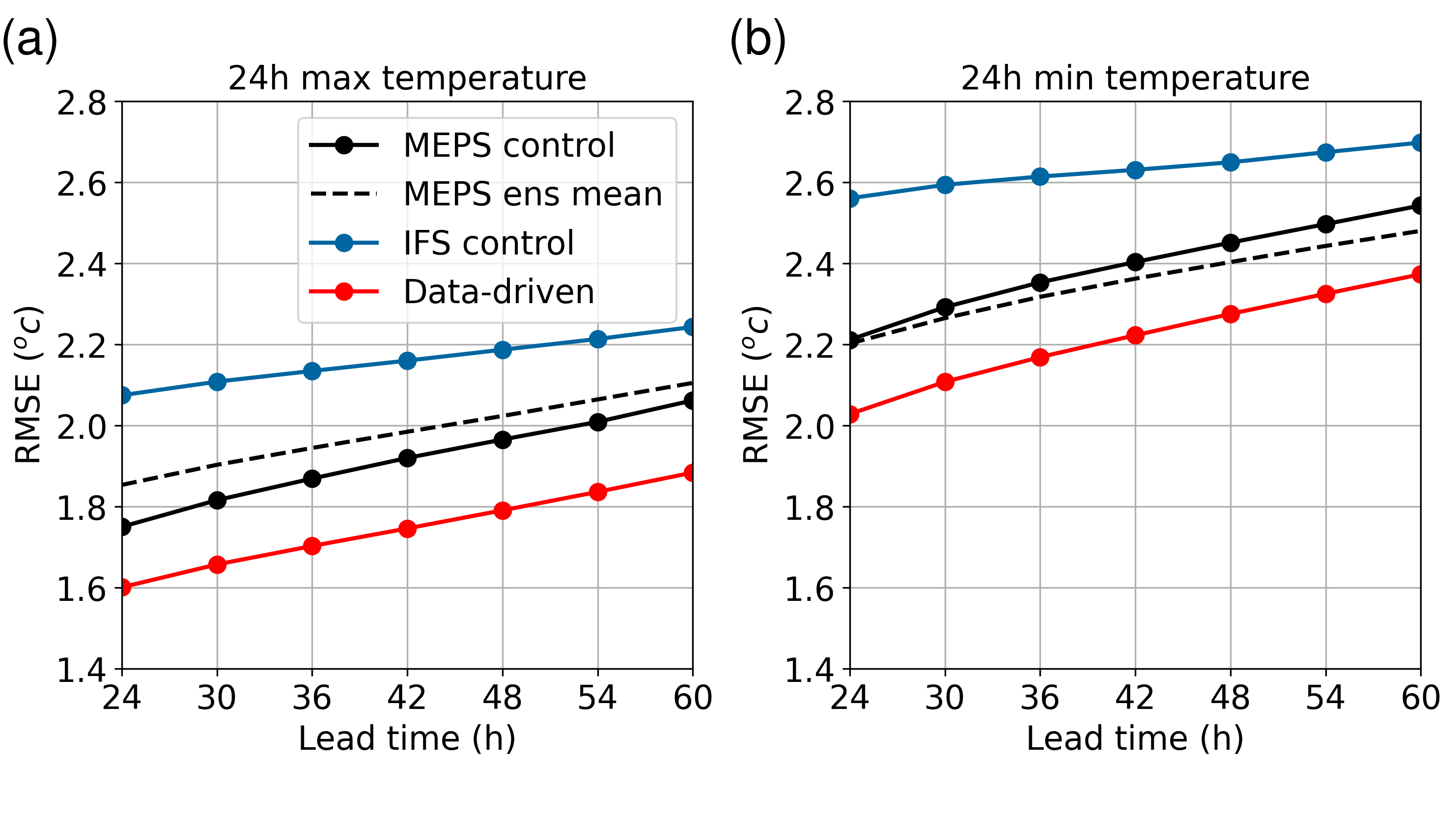}
    \caption{Root mean square error for \qty{24}{h} maximum temperature (a), and \qty{24}{h} minimum temperature (b).}
    \label{fig:t2m}
\end{figure}

The \acrshort{ddm} underestimates the standard deviation of temperatures across a 24-hour period (not shown). The consequence of this is that daily minimum temperatures are overestimated and daily maximum temperatures are underestimated. Despite these biases, the \acrshort{ddm} outperforms \acrshort{meps} in RMSE for both aggregated variables (\cref{fig:t2m}a--b).

\subsection{Wind speed}

To assess the model's ability to predict extreme events, we use the \acrfull{ets}, also known as the Gilbert skill score. For a particular wind speed threshold, the frequency of forecasts and observations exceeding or not exceeding the threshold are computed, thereby classifying events into hits ($a$), false alarms ($b$), missed events ($c$), and correct rejections ($d$). The \acrshort{ets} for a particular wind speed threshold is defined by:
\begin{equation}
   \text{ETS} = \frac{a-\hat{a}}{a+b+c-\hat{a}},
\end{equation}
where
\begin{equation}
   \hat{a} = \frac{(a+b)(a + c)}{a+b+c+d} ;
\end{equation}
The \acrshort{ets} penalises forecasts with a high number of false alarms and event misses.

\begin{figure}
    \centering
    \includegraphics[width=0.5\textwidth]{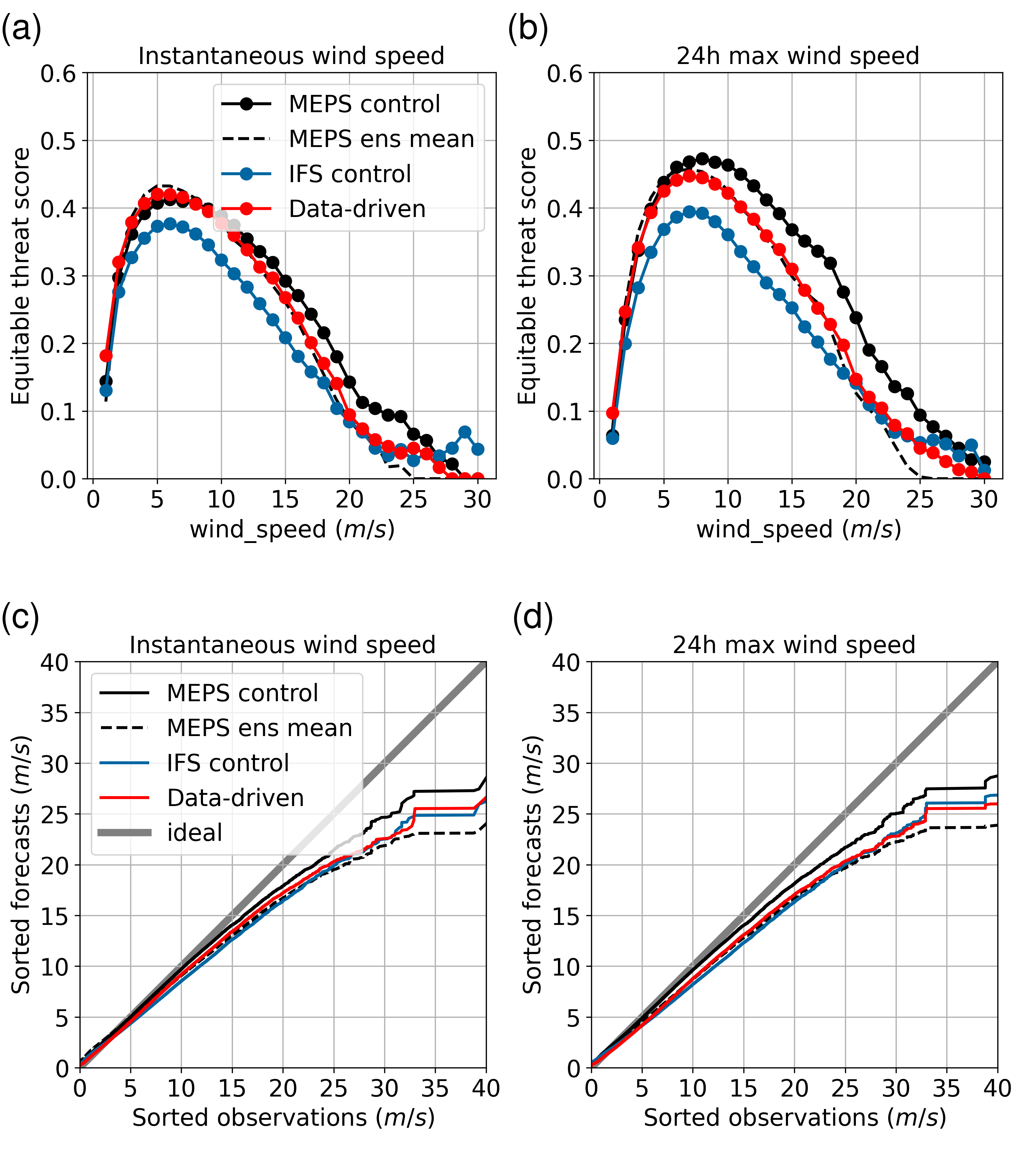}
    \caption{Wind speed assessment for lead time +60h. Equitable threat scores for various wind speeds are depicted for both instantaneous and \qty{24}{h} maximum wind speed in panels (a) and (b), respectively. Panels (c) and (d) are quantile-quantile plots.}
    \label{fig:ws10m}
\end{figure}

Spatial smoothing is significant for wind speed because of the double-penalty problem associated with \acrshort{mse} loss. When a model predicts an event slightly off its actual position, it gets penalized twice: once for predicting the event at the incorrect location, and again for not predicting the event at the correct location. Smoothing the wind field improves the \acrshort{rmse} but decreases the forecast's effectiveness in warning against strong wind events. The \acrshort{ddm} outperforms the IFS in terms of ETS for most thresholds, which is largely due to IFS's poor ability to represent strong winds in the Norwegian mountains. It also has ETS scores close to \acrshort{meps}, except for wind speeds above \qty{15}{m/s} (\cref{fig:ws10m}a). We also see that the \acrshort{ddm} produces fewer strong wind events than what is observed (\cref{fig:ws10m}c). This is problematic for public weather forecasting. A simple solution is to adjust the distribution of winds to better match the distribution of climatology by a separate post-processing step.

For weather warnings, the timing of an event is often less important. To evaluate this need, we computed the \qty{24}{h} maximum wind speed (from the four instantaneous values within the time period). The \acrshort{meps} control outperforms the \acrshort{ddm} for an even greater range of wind speed thresholds (\cref{fig:ws10m}b) than for instantaneous wind speed. This is because the \acrshort{ddm} has an increasing number of misses for higher thresholds, and has skill closer to the ensemble mean.

\subsection{Precipitation}

\begin{figure}
    \centering
    \includegraphics[width=0.5\textwidth]{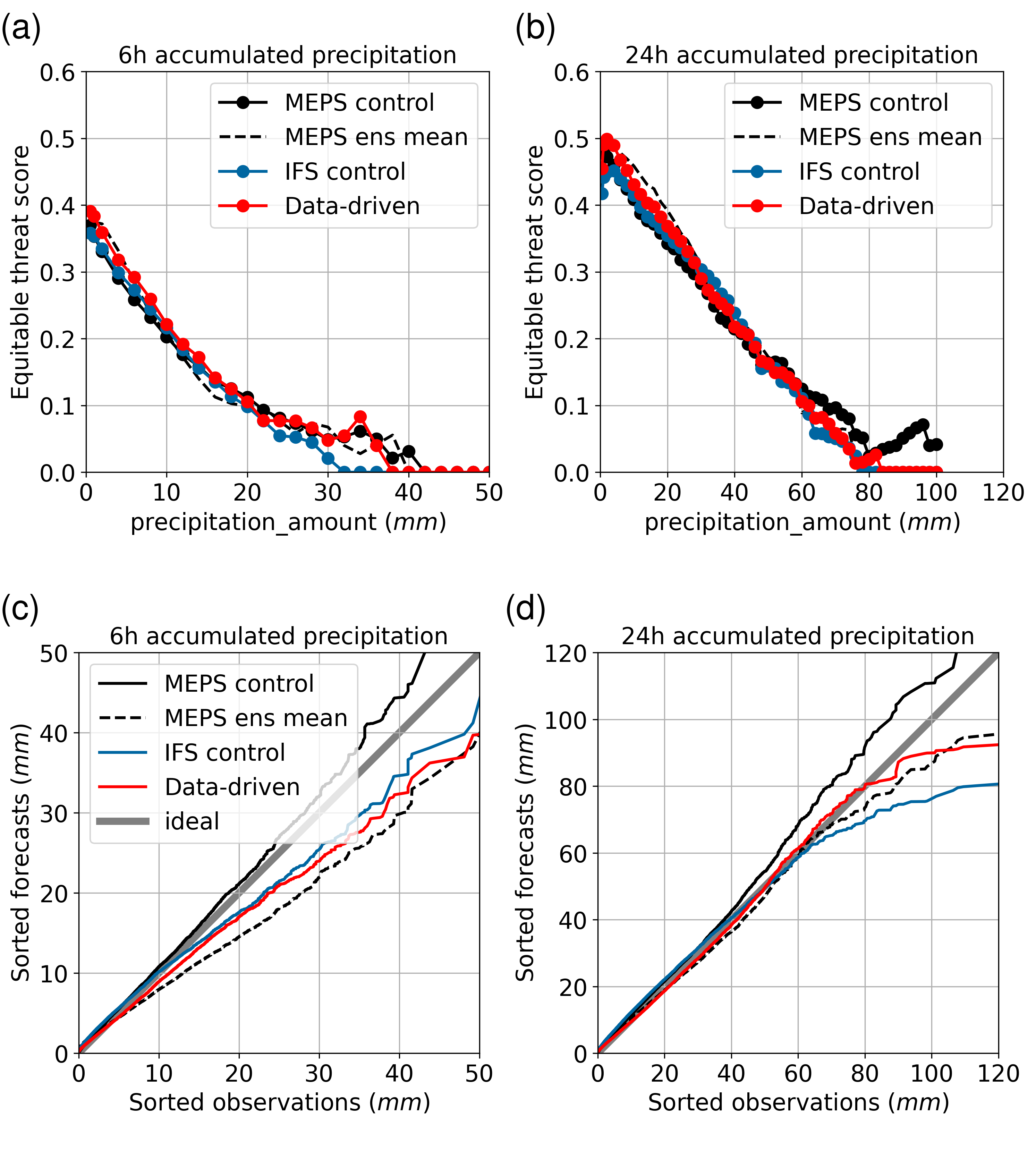}
    \caption{Precipitation assessment for lead time +66h. The panels are organized similar to \cref{fig:ws10m}.}
    \label{fig:precip}
\end{figure}

For \qty{6}{h} accumulated precipitation, the \acrshort{ddm} performs similar with respect to ETS to the NWP baseline models (\cref{fig:precip}a). As with wind speed, we see that the \acrshort{ddm} underestimates higher precipitation amounts (\cref{fig:precip}c), which can be attributed to spatial smoothing.

Many of our users are interested in aggregated precipitation over longer periods, particularly at a daily timescale. Although the \acrshort{ddm} is not directly trained to produce accurate \qty{24}{h} precipitation accumulations, we see that the model is competitive compared to the baseline models (\cref{fig:precip}b). The underestimation of extreme precipitation is less evident at the \qty{24}{h} timescale (\cref{fig:precip}d) than at the \qty{6}{h} time scale, and is likely due to the fact that any temporal smoothing at the \qty{6}{h} time scale does not negatively affect the \qty{24}{h} aggregated values.

\begin{figure}
    \centering
    \includegraphics[width=0.5\textwidth]{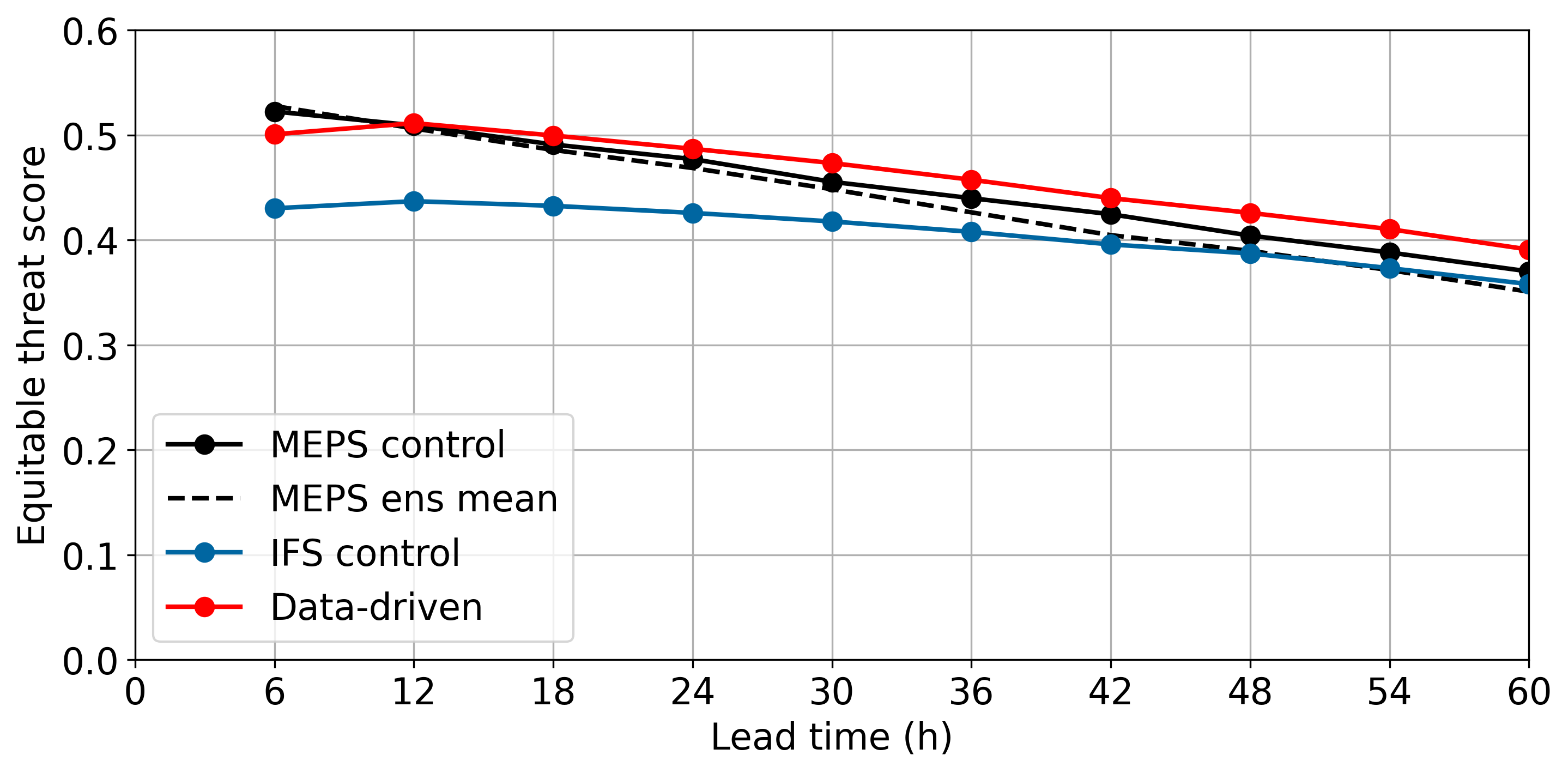}
    \caption{Equitable threat score for \qty{6}{h} precipitation exceeding \qty{0.5}{mm} for different lead times.}
    \label{fig:rain_no_rain}
\end{figure}

For public weather forecasting, discriminating between a dry period and a rainy period is important as this affects the weather symbol we use to represent the period. For this, we use a threshold of \qty{0.5}{mm}, above which rain will appear in our weather symbol. \cref{fig:rain_no_rain} shows that the \acrshort{ddm} is better able to discriminate between the occurrence and non-occurrence of precipitation than the NWP models for lead times beyond \qty{6}{h}.

\section{Conclusions and future work}
\label{sec:conclusions}

We have shown that a stretched-grid approach to regional data-driven modelling outperforms a state-of-the-art high-resolution NWP model for certain key parameters used in public weather forecasting, including instantaneous \qty{2}{m} temperature, \qty{24}{h} minimum and maximum temperature, and whether \qty{6}{h} precipitation exceeds \qty{0.5}{mm}. The model also provides competitive forecasts of instantaneous \qty{10}{m} wind speed, \qty{24}{h} maximum wind speed, and \qty{6}{h} and \qty{24}{h} accumulated precipitation, though the model tends to underestimate extremes.

The performance of the \acrshort{ddm} is promising enough that it warrants use in public weather forecasting. However, several challenges must be overcome before we can use the model operationally. Firstly, forecasts at an hourly time scale is needed to serve the needs of our users. This will pose an even bigger challenge regarding GPU memory and it is still an open question if such a model can perform as well as a model trained with a \qty{6}{h} time step. Alternatives to auto-regressive approaches could be considered, such as training a separate model that interpolates 6-hour scenarios in time.

A second challenge is the need to provide probabilistic forecast \cite{price2023,lang2024a} parameters, such as the quantiles of the probability distribution, which we use in our app to indicate the risk of heavy rainfall and strong winds. This can be solved by training an ensemble model or by directly modelling the quantiles using an appropriate loss function for scoring quantile forecasts.

To further improve the model in general, using more observational data could be important. For temperature, the \acrshort{ddm} was able to utilize the information from assimilated observations better than the NWP model. Additional observations could be assimilated into the input grid before training, but a more general approach would be to use them directly with a separate encoder. Using crowdsourced observations, which have previously been shown to add value in operational weather prediction \citep{nipen2018}, is an avenue we will explore further.

Due to their low computational requirements, \acrshort{ddms} will allow us to run models for longer lead times than is typically affordable with high-resolution \acrshort{nwp} models. This potentially allows us to use a single model for a range of timescales, instead of merging model runs from short-range, medium-range, and extended-range NWP systems as we do today. Operationally, these systems require separate post-processing as the models have different resolution and biases. End users would benefit from seamless forecasts without noticeable jumps across timescales.




\subsection*{Acknowledgments}
This work was supported by computing and storage resources provided by Sigma2 -- the National Infrastructure for High-Performance Computing and Data Storage in Norway (Grant No. NN10090K).

\subsection*{Data availability statement}
The verification data (observations and forecasts) in this article are made accessible in Verif-format \citep{nipen2023} on Zenodo (\url{https://zenodo.org/communities/verif/}).


\bibliographystyle{ieeetr}
\bibliography{references.bib}  



\end{document}